\documentclass[10pt]{article}
\usepackage{amsmath,bm}
\usepackage{graphicx,pdflscape}
\usepackage[toc,page]{appendix}

\usepackage{color,subfigure}
\usepackage{soul}
\graphicspath{{images/}}
\usepackage{mathtools}
\usepackage{amsmath}
\usepackage{hyperref}
\hypersetup{
    colorlinks,
    citecolor=red,
    filecolor=cyan,
    linkcolor=blue,
   urlcolor=magenta
 }

\graphicspath{{images/}{../}}
\usepackage{epstopdf}

\usepackage{tabularx}

\usepackage{capt-of}

\usepackage{listings}

\usepackage{soul}

\definecolor{dkgreen}{rgb}{0,0.6,0}
\definecolor{gray}{rgb}{0.5,0.5,0.5}
\definecolor{mauve}{rgb}{0.58,0,0.82}

\title{ Super Lax Pairs  and Infinite Symmetries in The $1/r^2$ System}
\author{B. Sriram Shastry\\
AT \& T Bell Laboratories\\
Murray Hill, N.J.,07974\\
and\\
Bill Sutherland\\
Department of Physics,University of Utah\\
Salt Lake City, Ut., 84112}
\date{ 17 November 1992}

\begin{document}
\maketitle

\begin{abstract}

We present an algebraic structure that provides an interesting
and novel link
between supersymmetry and quantum integrability.
This structure underlies two  classes of models that are exactly solvable
in 1-dimension and belong to the $1/r^2 $ family of interactions. The algebra
consists of the commutation between a ``Super- Hamiltonian'', and two
other operators, in
a Hilbert space that is an enlargement of the original one by
introducing fermions. The commutation relations reduce to quantal Ordered
Lax equations when projected to the original subspace,
and to a statement about
the ``Harmonic Lattice Potential''  structure of the Lax operator.
These  in turn lead to a highly automatic proof of the integrability
of these models.  In the case of the discrete $SU(n)-1/r^2$ model, the
`` Super-Hamiltonian'' is again an $SU(m)-1/r^2$ model with a related $m$,
providing an interesting hierarchy of models.

\end{abstract}

\newpage

$\bullet${\bf Introduction:} In this work we present a novel algebraic
structure
that has arisen, in the course of our
recent work on the $1/r^2$ family of many body problems. This structure seems
to be of possible interest  in several many body systems, and also in the
context of field theoretic problems   that lead to
a study
of matrix models of different kinds, including those with fermionic
degrees of freedom \cite{hep}. We find a new and
intimate relationship, within the models considered here,
between concepts that are of great interest in
their own right, namely supersymmetry \cite{susy}
 and quantum integrability.  We find,
remarkably enough,
that the quantum integrability of certain bosonic quantum systems is easier
to understand by  enlarging the Hilbert space, and
embedding these in
problems  containing fermionic
degrees of freedom as well. Under this enlargement,
certain non trivial Lax relations for the bosonic systems, turn into
simple commutators in the
enlarged system. The  Lax equations, unlike in the classical
case \cite{moser,crm,com1}, do not imply integrability in the quantum
 models in general due
to severe quantum ordering problems. In the
models considered here, however, integrabilty follows
from the structure of our equations in a highly automatic fashion.
 From the point of view of supersymmetry,   nontrivial
models exhibiting  this symmetry, are shown to be made up of
more fundamental
operators, which are invisible at the usual
level of description.

\qquad Our  examples in this work are the
continuum Sutherland- Calogero- Moser (SCM) $1/r^2$ \cite{scm,moser} system,
and the
discrete $SU(n)$ symmetric $1/r^2$ system solved by Shastry \cite{bss1}
and independently by Haldane \cite{haldane}. We first present the algebraic
structure,
and show how quantum integrability follows simply from this. We next discuss
the SCM
system where the algebraic structure is realized, and also
where supersymmetry in the usual sense \cite{susy} is
 fulfilled.
We next study the discrete  $SU(n)-1/r^2$   mode, this system
has received considerable attention very recently from
 diverse points of view
 \cite{bss2,fowler,kawakami,kuramoto,gebhard,haldfrench,kitawa}.
 In this work, we show that this model
fits  very naturally into the above structure.
Morover, our  scheme uncovers a fascinating hierarchical structure
of Hamiltonians,
wherein at each stage, the larger
Hilbert space ``Super-Hamiltonian'', has again the
form of the $SU(n)- 1/r^2$ model, but with a different number of components,
some
necessarily fermionic.

$\bullet${\bf General Algebraic Structure:}
We begin by highlighting the general algebraic structure that
emerges from the   detailed models, requiring a triad of operators,
$\mu$,
${\cal H} $ and $ {\cal L} $  with specific commutation relations. We
consider a system of N bosonic degrees of freedom\cite{com2},  say N bosons or
N
sites in a spin chain, and append to this set N fermionic degrees of
freedom,
i.e. operators $c_n, c_m^{{\dag}}$ obeying canonical anticommutation rules.
The ``uniform mode'' operator $\mu= \sum_{n=1}^N c_n$, an unnormalized
fermi operator ( $\mu^2=0$ ),  is the first member of the triad. We next
require a Hermitean  Super-Hamiltonian ${\cal H}$ of the form
\begin{equation}
{\cal H}= H_b+ H_f,
\end{equation}
with $H_b$ consisting purely of bosonic variables
and $H_f= \sum_{i,j} M_{i,j} c^{{\dag}}_i c_j$ , with $M_{i,j}$
purely bosonic. The third operator is the Super Lax operator ${\cal L}$ in
the form
 \begin{equation} {\cal L}= \sum_{i,j} L_{i,j} c^{{\dag}}_i c_j, \end{equation}
 where again
$L_{i,j}$ is bosonic
and obeys $L_{i,j}^{{\dag}}=L_{j,i}$ so that ${\cal L}$ is Hermitean. The two
fundamental commutations we require are
\begin{equation}
[{\cal H},{\cal L}]=[{\cal H},\mu]=0.   \label{fund}
\end{equation}
The commutation with $\mu$ requires a constraint on the form of $M$
namely $\sum_i M_{i,j}=0=\sum_j M_{i,j}$. The Super Lax relation
requires a highly nontrivial constraint on the functional forms of the
operators $L, M$ and  $H_b$.
The operators ${\cal L} $ and $\mu$ do not commute with each other in
general.
It follows from  Eq(\ref{fund}) that any operator function of $\mu, {\cal L}$
say $f(\mu,{\cal L})$ commutes with ${\cal H}$. This scheme provides us with an
elegant formulation of the integrability of the purely bosonic model
$H_b$.  This requires us to introduce the notion of bosonic projection
of an arbitrary ``super ''operator: $f \longrightarrow
\hat{f}_{\alpha,\beta}= \; <0|\mu^\alpha f
(\mu^{{\dag}})^\beta |0>$, where $|0>$ is the fermionic vaccuum state defined
by $c_n |0> =0$,
with $\alpha,\beta=0,1$ giving four possible
results. It is easy to see that
\begin{equation}
\{ [{\cal H},f]=0 \}\;\Rightarrow [H_b,\hat{f}_{\alpha,\beta}]=0.
\label{projcom}
\end{equation}
To see this take the matrix element of $[{\cal H},f]=0$ in
$(\mu^{{\dag}})^\beta
|0>$ and
$(\mu^{{\dag}})^\alpha
|0>$,
  giving $[H_b,\hat{f}_{\alpha,\beta}]=<0|(\mu)^\alpha [f, H_f]
(\mu^{{\dag}})^\beta|0>$, however
Eq(\ref{fund}) allows us to commute $H_f$ past $\mu$ and annihilate
$|0>$ proving Eq(\ref{projcom}). In practice, only one of the four
projections is of use in giving non trivial results.  The scheme for
constructing an infinite number of commuting operators is clear, we take
various functions $f$ and project them. Thus the integrability of $H_b$
is closely connected to the above embedding into our algebra
Eq(\ref{fund}). Specific examples of functions are easy to find, one
obvious choice is to let $f=L^n$, where n is an integer. In this case
the useful projection is $\hat{({\cal L}^n)}_{1,1}=\sum L_{i_1,i_2} L_{i_2,i_3}
\ldots L_{i_n,i_{n+1}}$, which may be written simply as $Tr(L^n
\Lambda)$ viewing $L$ as a $N \times N$  matrix and the special matrix
$\Lambda_{i,j}=1 \; \; \forall \; i,j$. Another interesting class of
functions is ${\cal G}(t)= \{\mu(t),\mu(0)^{{\dag}} \}$, where $\mu(t)=\exp(i
t)\;\mu
\;\exp(-it)$ and $t$ is a spectral parameter. Expanding in $t$ we find
$\mu(t)$ contains $1,3,5 \ldots$ fermi operators and hence ${\cal G}(t)$
contains even fermi operators. Explicit examples are detailed below.

\qquad The first of the fundamental relations Eq(\ref{fund}), is in
fact one in the sense of Lax and Moser \cite{moser}. In the examples considered
in
this work, the commutator
$[{\cal H},
{\cal L}]$
reduces, remarkably enough, to a bilinear in fermions with a bosonic
coefficient, the vanishing of this coefficient is a quantal Ordered Lax (OL)
Equation
\begin{equation}
[L_{i,j},H_b]= \sum_k (M_{i,k} L_{k,j}- L_{i,k} M_{k,j}). \label{lax}
\end{equation}
In this equation $L$ and  $M$ are bosonic operators, and so the ordering of
the terms in Eq(\ref{lax}) is crucial. The early work of Calogero,
Ragnisco and Marchioro (CRM) \cite{crm} quantized the classical Lax equation
written down by Moser \cite{moser}, by antisymmetrizing the RHS of
Eq(\ref{lax}) in
the quantum sense. The above Ordered Lax (OL) equation has a natural matrix
product like  order built into it
leading to a ``telescopic'' cancellation of internal terms, whereby
$[L^n,H_b]=(M.L^n-L^n.M)$. If further the matrix relation $M.\Lambda=
\Lambda.M=0$, with $\Lambda_{i,j}=1$, then we see that
$\Lambda.[L^n,H_b]. \Lambda=0$. The condition $M.\Lambda=0$ is
recognizable as the condition $[H_f,\mu]=0$. We emphasize that in the
present treatment the OL Equation(\ref{lax}) does not lead to
integrabilty, unlike in the classical Lax case, we further need the
condition $M.\Lambda=0$.

$\bullet$ {\bf Continuum $1/r^2$ model:}
 We begin with the SCM model \cite{scm}, where ($x_i \in [0,L]$)
\begin{equation}
H_b=\sum_i p_i^2+ 2 \lambda(\lambda-1)\sum_{i < j}
\sin^{-2}(\phi(x_i-x_j)) -E_0 \label{billh},
\end{equation}
with $\phi= \pi/L$ and $E_0=N(N^2-1)\phi^2 \lambda^2/3$, and write for
the Lax operator
\begin{equation}
L_{i,j}= p_i \; \delta_{i,j} +  \chi_{i,j} \; (1-\delta_{i,j})
\label{billlax}
\end{equation}
where the function $\chi$ is to be determined. The commutator $[{\cal H},{\cal
L}]$ would be
trivially bilinear in the $c's$ if the $L$ and $M$ commuted,
independently of the statistics of the variables $c's$. As it is,
we have all the off diagonal elements of $L$ commuting with $M's$ (
 assumed to be  functions of $x_i$ ), and  only
$L_{i,i}=p_i$, does not commute with $M's$. We note however that in the
case of $L_{i,i}$ the four $c's$ operator generated is $c_i^{{\dag}} c_i
c_i^{{\dag}} c_k$, which reduces to $c_i^{{\dag}} c_k$, if we make the choice
that $c's$ are fermions. This choice of the statistics of $c's$ enables
the commutator to be reduced entirely to a bilinear in the fermions:
\begin{equation}
[{\cal L},{\cal H}]=\sum_{i,j} c^{{\dag}}_i c_j \{[L_{i,j},H_b]+\sum_k(L_{i,k}
M_{k,j}-M_{i,k} L_{k,j})\} \label{superlax}
\end{equation}
The condition for its vanishing is just the OL Equation
Eq(\ref{lax}). The functions $\chi$ and $M$ can be found
by writing $M_{i,j}=\delta_{i,j} \sum_k d_{i,k}+(1-\delta_{i,j})m_{i,j}$,
and writing the interaction more generally in Eq(\ref{billh}) as $1/2\sum
v_{i,j}$,
giving functional relations $m_{i,j}=2 \chi'_{i,j}$, $v_{i,k}= (const
-d_{i,k}+2 \chi_{i,k}^2)$ and  $\chi_{i,j}(d_{i,k}-d_{j,k})=\chi_{i,k}
m_{k,j}-m_{i,k} \chi_{k,j}$.  This set of
equations has a large class of solutions,
 with elliptic functions  for
$v_{i,j}$, being the most general answer. However, in view of Eq(\ref{fund}),
we demand $d_{i,k}= - m_{i,k}$, which cuts down on the allowed solutions
drastically. The essentially unique solution (with periodic boundaries) is
\begin{equation}
\chi_{i,j}=\phi \lambda \cot(\phi(x_i-x_j))
\end{equation}
\begin{equation}
M_{i,j}=2 \lambda \phi^2[ \delta_{i,j} \sum_k' \sin^{-2}\phi(x_i-x_k)-
(1-\delta_{i,j}) \sin^{-2} \phi(x_i-x_j) ] \label{billm}.
\end{equation}
Another, essentially equivalent
solution is found by replacing $\lambda  \longrightarrow (1-\lambda)$
in the above Eq(\ref{billm}).
Note that $\sum_i M_{i,j}=0$ in this case by the ``harmonic
lattice potential'' structure of the operator M leading to the matrix
equation $M.\Lambda=0=\Lambda.M$.
We remark that the structure of $M $ and $L$ is identical to that  found in the
classical case by Moser, and also
by CRM.

\qquad In this model we have also supersymmetry in the usual sense
\cite{susy} (n=1
complex SUSY): we define a new fermi like operator
\begin{equation}
\zeta=\sum Q_i c^{{\dag}}_i;  \; \; Q_i \equiv \sum_j L_{i,j} \label{zeta}
\end{equation}
with $\zeta=[{\cal L},\mu^{{\dag}}]$. The operators $[Q_i,Q_j]=0$, whereby
$\zeta^2=0$. In fact the ground state of Eq(\ref{billh}) is annihilated
by each of the $Q_i$.
 We can use the standard  supersymmetry construction and form the
operator
\begin{equation}
H_{ss}= \{\zeta,\zeta^{{\dag}} \}.   \label{hss}
\end{equation}
By construction we have the property $[H_{ss},\zeta]=0$.
 Supersymmetry of
the system in this sense was already noted in \cite{bss2}.
It is easy to show by explicit calculation for this model
that $H_{ss}= {\cal H}$.
Note that {\bf all} the operators in this supersymmetric theory
are generated by two underlying fundamental operators ${\cal L}$ and $\mu$.

\qquad We remark that the entire structure presented here goes through
for the case of open boundaries, i.e. the Calogero system. The results
may be obtained by similar calculations, or most simply by
making replacements in the  operators:  $[\phi \cot(\phi x)] \rightarrow
1/x$ and $[\phi/ \sin(\phi x)] \rightarrow 1/x$.

$\bullet${\bf Discrete $ SU(n)-  1/r^2$ model:} Here we consider
the discrete $1/r^2$ model
with
\begin{equation}
H_b= \frac{1}{2} \sum_{i,j} v_{i,j} P_{i,j} \label{hbss}
\end{equation}
where $v_{i,j}=v_{j,i}$ is the interaction. The exact
solution \cite{bss1,haldane} was given
for the case
$v_{i,j}=\phi^2/\sin^2(\phi(x_i-x_j))$, but
we shall keep its form general until later. Further the $x_i$ are
on a uniform lattice $x_i \in Z_N$.
Here $\phi=\pi/N$ and $P_{i,j}$ the permutation operator in $SU(n)$
(it is $\frac{1}{2}(1+\vec{\sigma_i}. \vec{\sigma_j})$ for spin 1/2 hardcore
bosons
corresponding to $n=2$). In this model we make the choice for the Lax operator
\begin{equation}
L_{i,j}= (1-\delta_{i,j})\; l_{i,j} \;(P_{i,j}+c) \label{dislax}
\end{equation}
where $l_{i,j}$ and $c$ are undetermined as yet, and for the operator
$M$ we assume
\begin{equation}
M_{i,j}= \delta_{i,j} \sum_k' g_{i,k} (P_{i,k}+d) -(1-\delta_{i,j})
g_{i,j}(P_{i,j}+d),
\label{dism}
\end{equation}
with  $d$ and the function $g_{i,k}$ are as yet undetermined. Carrying
out the commutator in Eq(\ref{fund}), we find
\begin{equation}
[{\cal L},{\cal H}]=\sum_{i,j} c^{{\dag}}_i c_j([L_{i,j},H_b]+\sum_k(L_{i,k}
M_{k,j}-L_{k,j} M_{i,k})) + \sum_{i,j,k,l}[L_{i,j},M_{k,l}]c^{{\dag}}_kc_l
c^{{\dag}}_i c_j. \label{horrideq}
\end{equation}
The last term generates a four fermion term with distinct $i,j,k$,
$$\sum_{i,j,k}'c^{{\dag}}_k c_k c^{{\dag}}_i c_j [P_{i,j},P_{i,k}](l_{i,j}
\{g_{k,j}-g_{k,i} \}
+l_{i,k} g_{k,j}-l_{k,j} g_{i,k} ),$$
which is required to vanish, yielding a functional
constraint (for unequal $i,j,k$)
\begin{equation}
(l_{i,j}-l_{i,k}) g_{k,j}= (l_{i,j}-l_{k,j}) g_{i,k}.   \label{fnleq}
\end{equation}
The third term in the rhs of Eq(\ref{horrideq})
becomes $c^{{\dag}}_i c_j g_{i,k}
(l_{i,j}+l_{k,j}) [P_{i,j},P_{i,k}]$, and after some rearrangement,
 becomes
identical to Eq(\ref{superlax}). The vanishing of the quadratic form yields
again the OL Equation(\ref{lax}). The OL Equation can be worked out
readily. Omitting the constants $c,d$ we see that the OL Equation
becomes
\begin{equation}
\sum_k'[ P_{i,j} P_{i,k} (l_{i,j}
(v_{i,k}-v_{j,k})+g_{i,k}l_{k,j}-l_{i,k}g_{k,j})+
P_{i,k}P_{i,j}(v_{j,k}-v_{i,k}+g_{j,k}-g_{i,k})l_{i,j}]=0.
\end{equation}
This yields  $g_{i,j}=g_{j,i}=-v_{i,j}$, and further the same
constraint as
 Eq(\ref{fnleq}) results. This equation can be written as
$l(x+y)=[l(x)v(y)-l(y)v(x)]/(v(y)-v(x)).$
It is clear that if $l(x)$ is a solution then so is $l+const.$ Non
trivial solutions can be found using the even-ness of $v$, by writing
$y=-x+\epsilon$, and expanding $v(y)$ around $ -x$.  To leading order in
$\epsilon$, we find $l(\epsilon)=[1/\epsilon] (l(x)-l(-x))v(x)/v'(x)$,
which implies that nontrivial solutions in $l$ are odd functions, and
further that $l(x) \propto v'(x)/v(x)$. We thus end up with a functional
equation satisfied by $v$ alone. With periodic boundaries, this has a
unique solution $v \propto 1/\sin^2(x)$. This also determines  the
function
$l_{i,j}= i p
\cot(\phi(x_i-x_j)) + q$, where $p,q$ are real constants.
The constants in Eq(\ref{dism}) can be found
after a tedious calculation as $d=\frac{2}{3} c$.

\qquad The Super-Hamiltonian  ${\cal H}$ can be written explicitly using the
above solution (with $c=0$) as
\begin{equation}
{\cal H}= \sum_{i < j} v_{i,j}P_{i,j}  \Xi_{i,j}, \label{superduper}
\end{equation}
where $\Xi_{i,j}=1-(c^{{\dag}}_i-c^{{\dag}}_j)(c_i-c_j).$
It is clear that under a particle hole transformation $c_i \rightarrow
c^{{\dag}}_i$, the spectrum of ${\cal H}$ is inverted, and in the ``half
filled''
sector $N_{F}=N/2$, the spectrum has inversion symmetry about zero.
Morover if $|\Psi>$ is an eigenfunction of ${\cal H}$ and $\hat{
N_F}$, with eigenvalues $E$ and $N_F$, then $\mu |\Psi>$ is either
null, or is degenerate with $|\Psi>$, with $N_F-1$ fermions.

\qquad It is a remarkable  and unexpected
fact that $\Xi_{i,j}$ also satisfies the generator relations for the
permutation group $\Pi_{i,j}^2=1$ and
$\Pi_{i,j}\Pi_{j,k}\Pi_{i,j}= \Pi_{j,k}\Pi_{i,j}\Pi_{j,k}$, and so does the
composite operator $\Xi P$. We thus see that the Super Hamiltonian is
again an $SU(m)-1/r^2$ model. The difference of course, is that the added
fermions
increase the number of components of the model. More explicitly, let the
starting ``bosonic'' model\cite{com2}  consist of $n_b$ bosonic species and
$n_f$ fermionic
species, so $n=n_b+n_f$.   We
always restrict ourselves to states in the Hilbert space where we have only one
particle at
each site.
By adding one new set of fermions to this model, we are forcing the fermions
to sit on top of one of the other particles. Once the new fermions are
``glued''
to pre-existing  particles, they hop along with
the carrier particles by the action of the Hamiltonian, and hence
we are effectively increasing the number of species. Clearly gluing a fermion
(boson)
to a boson or a fermion produces a fermion (boson)  or a boson (fermion)
and hence the Super-Hamiltonian is an  $SU(n')-1/r^2 $ system with $n'=2n$
and with $n'_f=n_f+n_b$ and with $n'_b=n_f+n_b$. The principal series
of models then is of the kind $SU(2^{\nu+1})$, with $2^\nu$ kinds of bosons
and an equal number of fermi species ($\nu=0,1, \ldots$). At the level $\nu$
the Super-Hamiltonian is related as in our algebraic scheme to
several different (lower $n$) models, in fact $2^\nu +1$ of them.
It should be clear, that the
model with added fermions has a spectrum that is highly
reducible, in the sense that it has several constants of motion, namely the
number of distinct  species, with regard to which the hamiltonian
breaks up into blocks.

\qquad  We have examined the  eigenspectrum of the ${\cal H}$ for small systems
numerically,
it shows fascinating
degeneracies and a rich structure, and we will return to its
discussion in a later work.

$\bullet${\bf Conclusions:} We have seen that the algebraic structure in
Eq(\ref{fund}) provides a simple and quite general framework in which to
understand the $1/r^2$ family of problems. The connection between Lax equations
supersymmetry and integrability is clarified in one set of problems.
 Further work is necessary
to understand the interrelations between various approaches
\cite{bss2,fowler,kawakami,kuramoto,gebhard,haldfrench,kitawa}, within this
unexpectedly rich family of problems.



\begin{thebibliography}{99}
\bibitem{com2} We warn the reader that the use of the term bosonic to describe
the degrees of freedom described in $H_b$ is slightly
misleading, since we can, and  occasionally do
consider the starting model to contain fermions as well as bosons.
\end{thebibliography}
\end{document}